\documentclass[journal,twoside]{IEEEtran}
\usepackage{cite}
\usepackage{amsmath,amssymb,amsfonts}
\usepackage{algorithmic}
\usepackage{graphicx}
\usepackage{subcaption}
\usepackage{textcomp}
\def\BibTeX{{\rm B\kern-.05em{\sc i\kern-.025em b}\kern-.08em
    T\kern-.1667em\lower.7ex\hbox{E}\kern-.125emX}}

\begin{document}

\title{Real-Time Small-Signal Security Assessment Using Graph Neural Networks}
\author{Glory~Justin,~\IEEEmembership{Student Member,~IEEE,}
 Santiago~Paternain,~\IEEEmembership{Member,~IEEE,}}

\maketitle

\begin{abstract}
Security assessment is one of the most crucial functions of a power system operator. However, growing complexity and unpredictability make this an increasingly complex and computationally difficult task. In recent times, machine learning methods have gained attention for their ability to handle complex modeling applications. Some methods proposed include deep learning using convolutional neural networks, decision trees, etc. While these methods generate promising results, most methods still require long training times and computational resources. This paper proposes a graph neural network (GNN) approach to the small-signal security assessment problem usimg data from Phasor Measurement Units (PMUs). Leveraging the inherently graphical structure of the power grid using GNNs, training times can be reduced and efficiency improved for real-time application. Also, using graph properties, optimal PMU placement is determined and the proposed method is shown to perform efficiently under partial observability with limited PMU data. Case studies with simulated data from the IEEE 68-bus system and the NPCC 140-bus system are used to verify the effectiveness of the proposed method. 
\end{abstract}


\section{Introduction}
\label{sec:introduction}
\IEEEPARstart{S}{ecurity} refers to the degree of risk a power system is able to withstand without interruption to power supply \cite{b1}. This risk refers to imminent disturbances or contingencies which occur unexpectedly throughout the day. Contingencies can range from three-phase faults on a transmission line to the loss of large components such as a transformer or generator. When a system does not have sufficient degree of security, it is exposed to severe system failures which can lead to cascading failures, widespread blackouts, economic losses and much more \cite{b1}. For example, in July 1997, a blackout occurred in New York City due to a personal mistake and an equipment outage, and it led to almost \$350 million in losses \cite{b2}. When a disturbance occurs, the power system is subjected to changes which can be small or large. It is important that the system is able to withstand these changes and settle to new operating conditions such that none of the system constraints are violated \cite{b3}. Thus security assessment analyzes the robustness of a power system to imminent disturbances.

Security assessment in power systems has many components to it such as static and dynamic security assessment, small-signal security assessment, transient security assessment, etc \cite{b3}. For small-signal security assessment (SSA), the focus is to predict the small-disturbance oscillator stability. This kind of stability is characterized by the system's damping of oscillations following a disturbance or change of system topology \cite{b38}. The disturbances are considered to be sufficiently small that linearized system equations can be used for analysis \cite{b3}. The eigenvalues of the linearized state matrix are then analyzed for sufficient damping. Historically, these computations were done in an offline planning environment. However, the the uncertainty of predicting future operating conditions in the power grid of today has led to a need for online security assessment \cite{b1}. The focus of this paper is online SSA in real-time.

Real-time SSA involves capturing the system current operating conditions and analysing for stability after a set of probable events \cite{b6}. This involves solving a series of time-domain simulations solving multiple differential and algebraic equations which take time. With larger networks, the number possible contingencies increases thus leading to more computational burdens \cite{b7}. The industry accepted time for a computational cycle is 15-30 minutes \cite{b6}. However, the use of conventional techniques generates outdated results due to the time taken for the computational cycle. Additionally, results are more relevant when the computational cycle is more frequent \cite{b6}. To reduce the computational burden and make computational cycles more frequent, a lot of research has gone into faster algorithms for online DSA. In recent years, machine learning algorithms are gaining attention for their ability to learn from data and the wide availability of data from modern monitoring systems \cite{b4}, \cite{b9}. In this paper, taking advantage of widely available data from PMUs, an efficient graph neural network approach is proposed which trains faster, performs efficiently in real-time and is more robust in the presence of missing data.

\subsection{Related Work}
Several methods have been proposed in the literature for security assessment problems. Some of these include direct methods such as transient energy function methods \cite{b4}, Lyapunov function approximations \cite{b10} and extended equal area criterion. However, these methods also require large amounts of computations like the traditional methods. These methods also in some cases, require detailed models of multi-machine power systems which are not easy to obtain \cite{b6}. With the availability of data, machine learning techniques are gaining more attention in the literature. According to \cite{b7}, machine learning for transient stability was first used in \cite{b11}. Here the authors use decision trees to analyse dynamic security. Since then, many other algorithms have been proposed for security assessment problems. In \cite{b12}, the authors use kernel ridge regression to determine the transient stability boundary for applications to dynamic security of a power system. In \cite{b13}, the authors propose an algorithm using support vector machines to predict transient stability after a fault from post-fault voltage variations at the generator buses. Some researchers also propose a combination of direct methods with machine learning such as in \cite{b6}. Here the authors use support vector machines to extract features from the transient energy function and determine the boundary between secure and insecure operating points.

More recently, deep learning has gained a lot of attention for solving complicated problems. Deep learning proposes to learn representations of data using multiple layers of nonlinear transformations \cite{b14}. While this concept has existed for decades, it was limited by training approaches, insufficient data, and computational power. However, with recent technological advancements, deep learning is gaining prominence with successful applications seen in image processing, speech recognition and object detection \cite{b7}, \cite{b14}. Leveraging these successes, research has been proposed to apply deep learning techniques to power system problems such as security assessment. In \cite{b14}, the authors use a deep autoencoder to extract important features for classifying stable and unstable operating points. In \cite{b31}, the authors also use a CNN to detect localized faults in real time. In \cite{b7} and \cite{b39}, the authors train a convolutional neural network (CNN) to classify images of power system snapshots as safe or unsafe. While these methods show promising results, they also require large amounts of data and long training times.   

For SSA, CNNs have been proposed such as in \cite{b7} and \cite{b39}. While both generate good results, they require large networks with millions of parameters and long training times. The CNN in \cite{b7} is trained using a network of 8 million parameters and trained for 24 hours. It also requires 1 million data points for training. In addition, the possibility of missing PMU data is not considered. In \cite{b39}, a feature extraction is used to reduce the CNN size to 563,489 and the possibility of missing data is also considered. However, the network size is still significantly large and thus requires significant computational resources both for offline training and online application. This paper proposes a GNN approach. Using GNNs, this network size can be significantly reduced thus reducing the computational resources and time needed both in offline training and online application. 

GNNs have previously been applied to power system problems such as power outage prediction \cite{b15}, fault-chain detection \cite{b32},  solving the optimal power flow problem \cite{b16} and power system state estimation \cite{b17}. Other authors have also proposed using GNNs for transient stability such as \cite{b40} and \cite{b41}  and for adaptive contingency screening \cite{b42}. The focus of this papaer is real-time SSA.  While other methods have been proposed for fast real-time analysis of small-signal stability such as \cite{b42} and \cite{b43}, most do not consider the impacts of missing data on the proposed method. They also still require significant computation for online application thus reducing real-time efficiency. Using GNNs, computational efficiency can be improved both in offline training and online testing. Also, since the GNN only looks at a few buses of crucial importance, it also shows a robustness to missing PMU data which is sometimes unavoidable.

GNNs have previously been proposed in literature for SSA. In \cite{b44}, the authors use recurrent graph convolutional networks (ResGCN) to predict the damping ratio and frequency for SSA. The authors in \cite{b45} also propose a graph-based temporal convolutional network to analyze for both small-signal stability and transient stability. While both proposed models show promising results, they do not show any advantage in efficiency, both for offline training and online testing. The effects of missing PMU data are also not analyzed. 

\subsection{Contributions of This Work}
This paper proposes a graph neural network approach to real-time SSA. While the autoencoder method and the CNN generate promising results, they still require large network sizes, long training times and large amounts of data to train. Using GNNs, the proposed method reduces the network size needed, reduces online testing time and improves training efficiency. The GNN also shows an added robustness to missing data. The major contributions of this paper are as follows:
\begin{itemize}
    \item A deep learning approach using GNNs is proposed for SSA. Using graph convolutions, the GNN is able to successfully extract a minimal amount of features, thus reducing the input size and network size needed for training. Thus the method proposed improves training efficiency, reduces the computational resources needed, and reduces the training time needed.
    \item Compared to conventional methods which require time-consuming simulations to obtain system dynamics, the proposed approach uses data which can be obtained from PMUs in real-time as model inputs. Without the need to compute multiple dynamic equations, the proposed method is shown to be more efficient for real-time application.
    \item Leveraging the graph properties of the power grid, optimal placement for PMUs is also proposed in case of limited PMU availability. The proposed method is then shown to perform efficiently under partial observability in conditions with missing PMU data .
\end{itemize}
The rest of this paper is structured as follows. Section \ref{sec_problem_des} provides more details about real-time SSA. Section \ref{sec_graph_neural_nets} provides an introduction to GNNs and the proposed method. Other than concluding remarks (Section \ref{sec_conclusion}) the paper finishes with  numerical experiments (Section \ref{sec_simulation_results}) on two different power grids, which showcase the advantage of the proposed method.

\section{Problem Description}\label{sec_problem_des}
The major objective of SSA is contingency analysis. The objective is to analyze the system and take preventive or corrective control actions such as rescheduling of active power generation, switching of Flexible AC transmission systems (FACTS), HVDC line MW transfer and load shedding \cite{b1}. For reliable operation, a power system has to be able to withstand credible contingencies. However, a power system cannot be built economically to withstand all contingencies \cite{b18} so a common criterion used for SSA is the N-1 contingency criterion \cite{b4}, \cite{b18}. This refers to the ability of an N-component power grid to withstand the loss of any single element either spontaneously or through a single, double or three phase fault \cite{b4}. To do this, the power system operator simulates a set of possible events individually and analyses the post-fault system for stability, as well as power transfer and thermal loading limits \cite{b18}. If the system is unstable or the limits are exceeded, the system is insecure and has to be re-dispatched. 

\begin{figure}    \centerline{\includegraphics[scale=0.6]{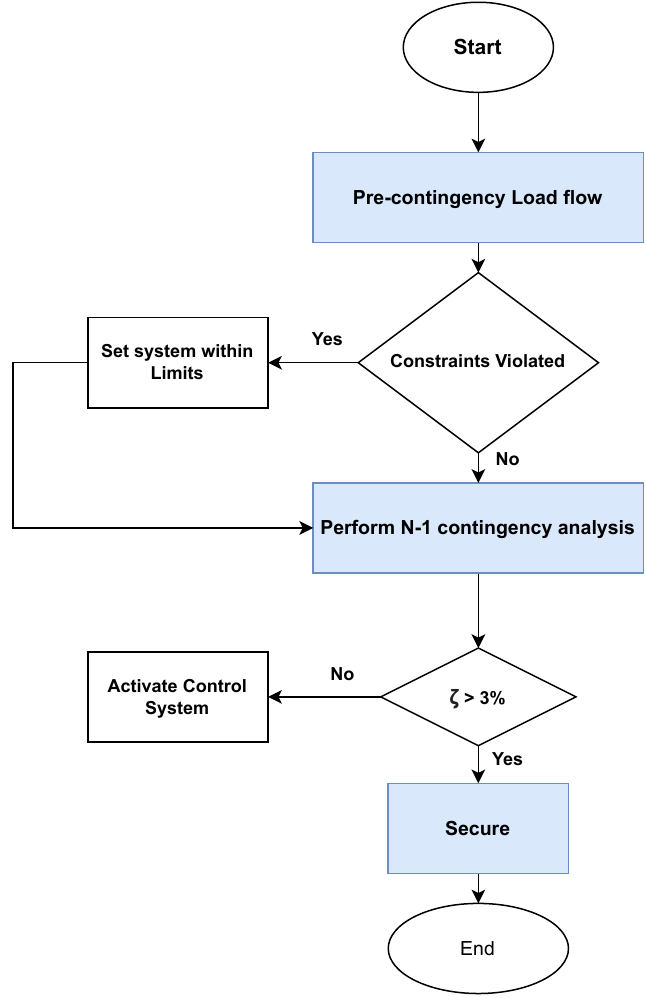}}
    \caption{A flowchart showing the steps involved for SSA using N-1 contingency analysis. The major computational burden is in performing the N-1 contingency analysis. This is where the proposed method is used.}
    \label{fig:flow_chart}
\end{figure}

As mentioned previously, SSA is directly related to small-signal stability analysis. In this analysis, the system variables are initialized and equations linearized around the initial values using the first-order Taylor expansion. In larger power systems, partial eigensolution methods are used. The oscillation modes are represented using the eigenvalues of the system matrices obtained from the systems dynamic equations \cite{b43}. The damping ratio $\zeta$ is a measure of the damping of an oscillation post-disturbance represented by a complex pole $\lambda = \sigma + j\omega$ . It is calculated using (\ref{eqn:damp_coeff})
\begin{equation}\label{eqn:damp_coeff}
    \zeta = \frac{-\sigma}{\sqrt{\sigma^2 + \omega^2}}.
\end{equation}
In this paper, we classify a system as safe if the damping ratio $\zeta$ is greater than 3\% as done in \cite{b7}. A flowchart showing the process for SSA using the N-1 contingency criterion is shown in Fig. \ref{fig:flow_chart}.

\begin{figure}
\centerline{\includegraphics[scale=0.8]{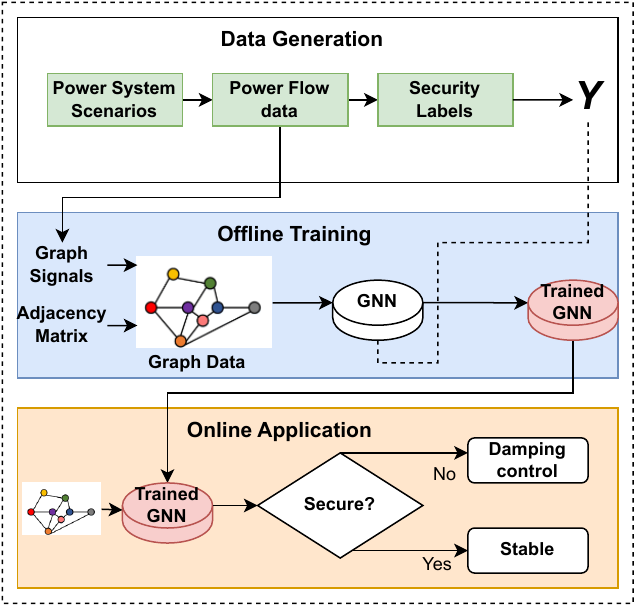}}      
\caption{Operation of the proposed method similar to the method in \cite{b44}. The Real-time SSA is carried out in 3 steps: first data generation, then offline training using graph data and lastly online testing using the trained GNN.}
\label{fig:GNN_operation}
\end{figure} 

On a 64-bit computer with Windows OS, 64 GB RAM and Intel core i9 proecessor,  each contingency takes 2.5 seconds on average. This amounts to needing $2.5\times86 $ seconds ($\approx 3.5$ minutes) to analyze for all 86 lines in the 68-bus system. For larger networks with hundreds or even thousands of buses, this compounds due to the increased number of possible contingencies. The time required for each contingency also increases due to increase in the number of equations and variables to compute. For the 140-bus system, each contigency takes 2.7 seconds on average and adds up to $2.7\times233 $ seconds ($\approx 10.5$ minutes) to analyze for all 233 lines. In this paper, instead of analyzing each contingency, a set of operating points are inputed to the GNN. This GNN is trained to classify operating points as safe or unsafe if safe for all the contingencies in consideration. 

The operation of the proposed method is shown in Fig. \ref{fig:GNN_operation}. First the data is generated from varying power system operating conditions. This data includes the post-fault power flow data including the bus voltages $V$, net active power $P$ and net reactive power $Q$ at each bus and active and reactive power flows at each line and the security labels for secure and unsecure operating points. This data is processed through the GNN and used to train the GNN to effectively classify secure and unsecure operating points.  The trained GNN is then used for real-time SSA. Using the GNN, the time needed for SSA can be reduced from minutes to less than a second, thus allowing instant security estimations.

\section{Graph Neural Networks}\label{sec_graph_neural_nets}
Graph neural networks (GNNs) are information processing architectures for signals supported on graphs \cite{b21}. Network data, such as the power grid, can be modeled as graph signals where the network topology is represented by data values assigned to nodes of the graph. This structure can then be exploited to successfully learn from network data  \cite{b22}. This results in GNNs having properties such as permutation equivariance, stability to perturbations and transferability \cite{b21}. These properties help to improve the generalization potential of GNNs over other deep-learning techniques for graph signals. For these reasons, GNNs have been successfully applied in recommender systems \cite{b33}, learning decentralized controllers \cite{b34} and learning optimal resource allocations in wireless communication networks \cite{b35}. Leveraging this success with networked data, we apply this principle to power system SSA, exploiting the structure of the power grid to improve efficicency and robustness.

GNNs can be described as analogous to CNNs, with graph filters replacing convolutional filters. CNNs exploit temporal or spatial convolutions to learn the data structure in temporal series and images, scale to large systems, and avoid overfitting \cite{b22}. While CNNs have shown success in data in regular domains e.g. images \cite{b36} and speech processing \cite{b37}, they perform poorly when learning from irregular network data \cite{b15}. In \cite{b7}, the authors apply CNNs to SSA with promising results. However, CNNs require large computational resources, with long training times and a large network with millions of parameters to learn satisfactorily. They also do not show much robustness to missing PMU data. Using GNNs, a smaller network can achieve the same results, with shorter training time and generalization potential, whereby a trained GNN can still perform satisfactorily on power grids with missing data at certain nodes.

\begin{figure}[htbp]
\centerline{\includegraphics[scale=0.4]{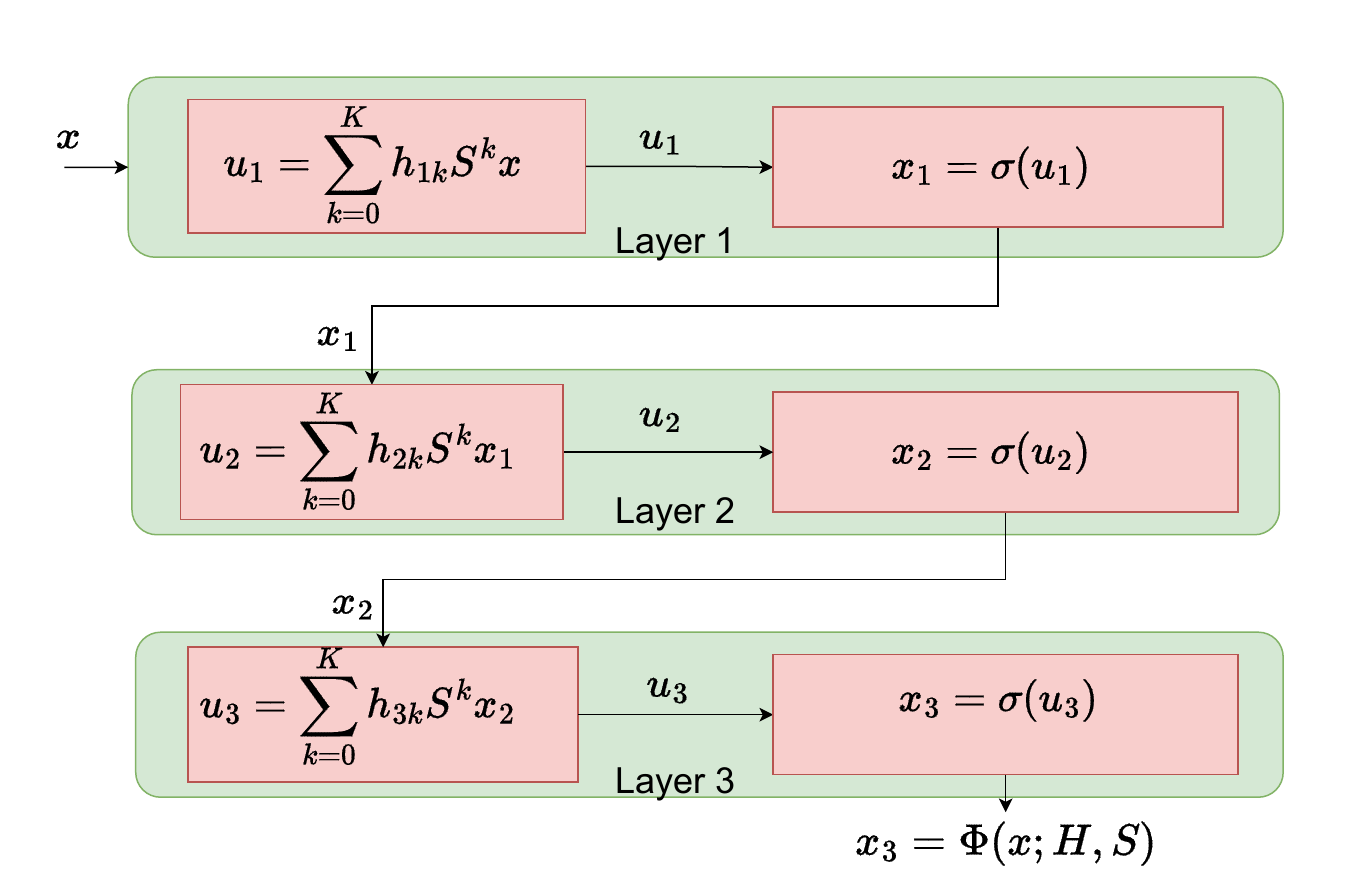}}
\caption{GNN formed from a stack of 3 graph perceptrons \cite{b20}}
\label{fig:GNN_stack}
\end{figure}

A graph filter is a polynomial in a matrix representation of a graph \cite{b21}. Consider an undirected graph $\mathcal{G} = (\mathcal{V,E,W})$ with a set of $N$ nodes $\mathcal{V}$, a set of edges $\mathcal{E} \subseteq\mathcal{V}\times \mathcal{V}$ and an edge weight function $\mathcal{W}: \mathcal{E}\rightarrow\mathbb{R}_{+}$. An $N \times N$ real symmetric matrix $S$, called the graph shift operator is associated with the graph. The $ij$ entry of the matrix is zero if the nodes $i,j\in\mathcal{V}$ are not connected and the entry is not zero if they are. Some common shift operators are the adjacency matrix and the Laplacian \cite{b19}. To describe the operating conditions of the system, we assign a graph signal $x\in\mathbb{R}^N$, where $[x]_i$ represents the signal at node $i$. The graph signal can be shifted over the nodes by using the shift operator $S$ so that 
\begin{equation}
    [Sx]_i = \sum_{j=1}^N[S]_{ij}[x]_j.
\end{equation}
Given a set of parameters $h = [h_0,\dots,h_k]^T,$ the graph convolution is
\begin{equation}
    H(S)x = \sum_{k=0}^Kh_kS^kx .
\end{equation}
The $k-$shifted signal contains a weighted sum of the $k-$hop neighbors. The graph convolution filters the graph signal $x$ with filter $H(S)$ which contains the filter weights $h_k$ to be learned during training. A pointwise nonlinearity $\sigma$ can be applied to the shifted graph signal to form a graph perceptron. $L$ Multiple perceptrons can then be stacked together to form the GNN such as the 3 layer network in Fig. \ref{fig:GNN_stack}. The output of layer $l\in L$ of the GNN is given by
\begin{equation}
    x_l = \sigma\left(H_l(S)x_{l-1}\right) = \sigma\left(\sum_{k=0}^Kh_{lk}S^kx_{l-1}\right).
\end{equation}
The optimal filter coefficients $H^*$ are learned by minimizing the loss function $\ell$
\begin{equation}
  H^* = \arg\min_H\frac{1}{M}\sum_{m=1}^M\ell\left(\Phi\left(x_m;H,S\right),y_m\right) 
\end{equation}
where $\ell$ is a loss that penalizes misclassifications in supervised learning e.g the cross entropy loss, $M$ is the number of samples and $\Phi\left(x_m;H,S\right)$ is the output from layer $L$~\cite{b16,b21,b22}.

For power grid SSA using GNNs, the graph $\mathcal{G} = (\mathcal{V,E,W})$ represents a power grid with a set of N buses $\mathcal{V}$ and a set of transmission lines $\mathcal{E}\subseteq\mathcal{V}\times\mathcal{V}$. The graph can be weighted or unweighted. However, for this problem using a weighted graph showed no significant changes in results so an unweighted graph is chosen, Since the graph is unweighted, the weight matrix $\mathcal{W}$ is the same as the adjacency matrix. The adjacency matrix serves as the graph shift operator $S$. $S_{ij}=1$ if there exists a line between buses $i$ and $j$ and $0$ otherwise. To describe the state of the grid, we assign 3 graph signals $x^f$ where $f=[1,2,3]$. These 3 graph signals are the active power $P$, reactive power $Q$ and voltage $Q$. For each node $i$ we have a vector $[X]_i = [v_i,p_i,q_i]$ $\in$ $\mathbb{R}^3$ where $v_i$ represents the voltage magnitude at bus $i$ , $p_i$ and $q_i$ represent the net active and reactive powers at bus $i$ respectively \cite{b7}. The net active and reactive powers are obtained by subtracting the load power at that bus from the generated power at the same bus. 

The GNN architecture used in this paper is the aggregation GNN first proposed in \cite{b23}. In this architecture, a single node is chosen and the graph signal is aggregated at this node. using successive communications between one-hop neighbours.  The aggregation sequence $z_i^f$ starts at the selected node $i$ where $f$ represents the states being considered, in this case voltage, active power and reactive power. The sequence is built by exchanging information with neighborhoods $S^kx^f$ and storing the result $[S^kx^f]_i$ at node $i$ for each $k$ from 0 to some number $K-1$. This results in the aggregation sequence 
\begin{equation}
  z_i^f = \left[[x^f]_i,[Sx^f]_i,\dots,[S^{K-1}x^f]_i\right]  
\end{equation} 
which can then be used to train a CNN \cite{b17}. For a 68 bus system for instance, the input data is reduced from 68 X 68 X 3 needed for the CNN to size $K \times 3$, where 3 is the number of power grid features in consideration. This reduced input data size requires less memory to train thus saving computational resources, training time and improving training efficiency. For improved performance, multiple nodes can also be combined for aggregation \cite{b24}. Using multiple nodes, the aggregation sequence becomes
\begin{equation}
  z_{i,j,\dots}^f=
\begin{bmatrix}
[x^f]_i&[Sx^f]_i&\dots&[S^{K-1}x^f]_i \\
[x^f]_j&[Sx^f]_j&\dots&[S^{K-1}x^f]_j\\
\vdots & \vdots & \ddots & \vdots 
\end{bmatrix}
\end{equation} 
for aggregation at nodes $[i,j,\dots].$

For the aggregation GNN, choice of node for aggregation is extremely important. This decision can be made based on various criteria such as the degree or using different centrality measures \cite{b24}. Other selection methods have also been proposed in \cite{b24} such as experimentally designed sampling and spectral proxies. For this paper, the aggregation node is chosen using degree and centrality measures, primarily the closeness centrality and eigenvector centrality. Node centrality is a measure of the importance of a node in a graph \cite{b27}. Closeness centrality $C_C(i)$ of a node $i$ in graph $\mathcal{G} = (\mathcal{V,E,W})$ looks at how quickly information is spread from that node to every other node in the network \cite{b27}. It is defined as the inverse of the sum of the shortest path lengths $s_G$ from node $i$ to every other node \begin{equation}\label{eqn:close_cen}
    C_C(i) = \left(\sum_{j\in V}s_G(i,j)\right)^{-1}.
\end{equation}
Eigenvector centrality however, depends on the neighbors of the node in consideration \cite{b27}. Nodes with neighbors of higher importance get higher centrality. For a node $i$ in graph $\mathcal{G} = (\mathcal{V,E,W})$, the eigenvector centrality $C_E$ is denoted by
\begin{equation}\label{eqn:eigen_cen}
    C_E(i) := \frac{1}{\rho}\sum_{(i,j)\in E} \mathcal{W}(i,j)C_E(j)
\end{equation}
for some constant $\rho$. More details on these centrality measures are shown in Section \ref{sec_choice_of_node}.

\section{Simulation and Results}\label{sec_simulation_results}
In this section the performance of the GNN is empirically evaluated. First the performance of the GNN and CNN are compared on efficiency, training time and accuracy with the IEEE 68-Bus system in Section \ref{sec:CNN_comparison}. Next the method for choice of aggregation nodes is described in Section \ref{sec_choice_of_node}. Multiple nodes are combined for multi-node aggregation for the larger 140-Bus system in Section \ref{sec_multi_node_agg}. The GNN is tested for robustness in limited PMU observability Section \ref{sec_transferability} and finally the testing time comparison is shown for the GNN in Section \ref{sec_testing_time}. 

\begin{figure}[htbp]
\centering
\includegraphics[scale=0.7]{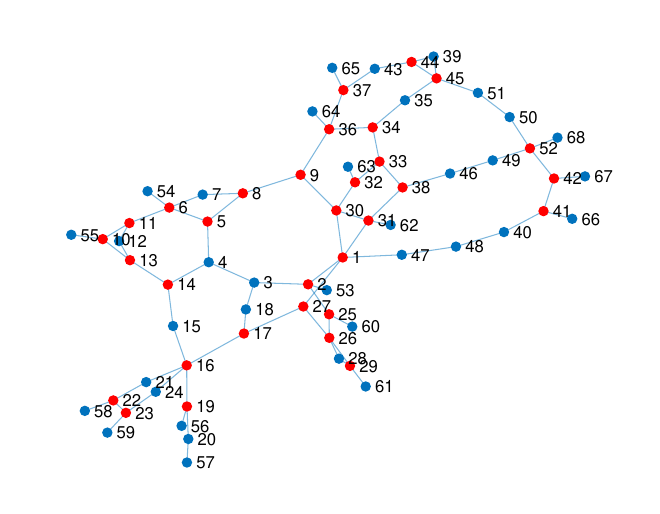}
\caption{IEEE 68-Bus system graph with 16 generators and 86 transmission lines \cite{b18}. Buses with highest centrality are highlighted in red.}
\label{fig:68bus}
\end{figure}

\begin{figure}[htbp]
\centering
\includegraphics[scale=0.6]{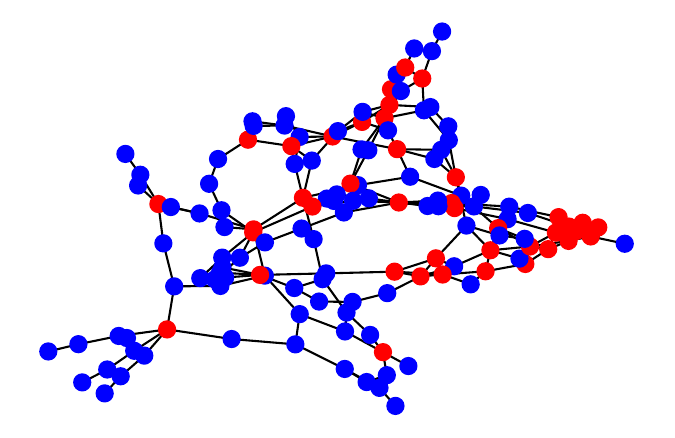}
\caption{NPCC 140-Bus system graph with 48 generators and 233 transmission lines \cite{b28}. Buses with highest centrality are highlighted in red.}
\label{fig:140bus}
\end{figure}

\subsection{Data Generation}\label{sec_data_generation}
The first step is the generation of data for training. In practice, this data would be obtained from PMU data, however, for this paper, simulated data is used. Two test systems are used for comparison with the GNN. The first is the IEEE 68 bus system which consists of 16 generators and 86 lines \cite{b18}. Fig.ure\ref{fig:68bus} shows a graph of the system with the buses of highest centrality highlighted. The second is the Northeast Power Coordinating Council (NPCC) 140 bus system \cite{b28}. Figure \ref{fig:140bus} shows a graph of the system with the buses of highest centrality highlighted. The initial data is for both systems is real data obtained from the sample problems in \cite{b18}. However, to generate multiple operating points, the load and generation profiles from the base case data are varied by a set of random numbers. For each of these load and generation profiles, first the optimal power flow is performed. Then, a set of N-1 contingencies $[C_1,C_2,\dots,C_n]$ consisting of a 3-phase fault on each of all transmission lines in the network is analysed. For the first system, this is $n=86$ contingencies and $n=233$ for the second test system. The data is generated using the Power System Toolbox \cite{b20} in MATLAB. A 3-phase fault is applied at time $t=0.1$s and cleared at $t=0.15$s. The oscillation modes are monitored after each fault and that operating condition is classified as safe if the damping for each of the 86 contingencies is greater than or equal to $3\%$. The bus voltages, net active power and net reactive power, as well as the line flows are then stored. These data points are used to train the GNN and a CNN for comparison.
Two datasets are used for this analysis and they are described below.
\begin{itemize}
    \item \textbf{Dataset 1}: For this dataset, data for the IEEE 68 bus system from \cite{b18} is used. The active power generation for each of the 16 generator buses is varied by a different randomly between 70-150\% of the base case. The load active and reactive powers at each of the 52 load buses are also varied randomly between 70-150\% of the base case. This range is chosen to give a more even distribution of safe and unsafe cases.  For each of these profiles, we perform N-1 contingency analysis as shown in the flow chart in Fig. \ref{fig:flow_chart}. First, the optimal power flow is performed and a fault is applied on each of the 86 lines in the system. The damping ratio for the oscillation modes after each of the 86 contingencies is considered and the operating point is classified as safe if the damping ratio $\zeta \geq 3\%$ for all 86 contingencies. Operating points where the power flow does not converge or system limits are violated are discarded. This generates 6253 data points with  56\% safe cases and 44\% unsafe cases. This dataset is used for comparison of the CNN and GNN on efficiency, training time and accuracy.
    \item \textbf{Dataset 2}: For this dataset, a larger sized network is used. The network used here is the NPCC 140 bus system \cite{b28}. This system has 48 generators, 140 buses and 233 lines. Like the the previous dataset, the active power generation, load active and reactive powers are varied randomly between 70-150\%. For each these profiles, the optimal power flow is performed and a fault is applied on each of the 233 lines in the system. The damping ratio for the oscillation modes after each of the 233 contingencies is considered and the operating point is classified as safe if the damping ratio $\zeta \geq 3\%$ for all 233 contingencies. Operating points where the power flow does not converge or system limits are violated are discarded. This generates 4725 data points with  29\% safe cases and 71\% unsafe cases.
\end{itemize}

\begin{figure}[htbp]\label{fig:CNN_arch}
\includegraphics[width=9cm, height=5cm]{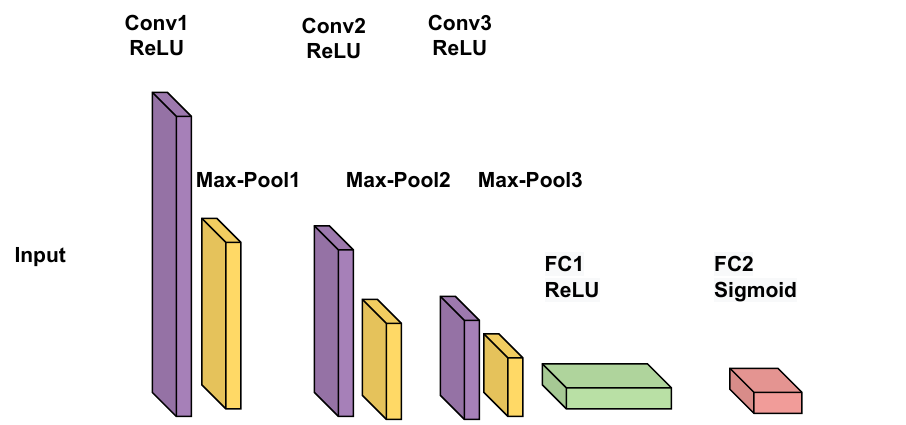}
\caption{CNN Architecture used for training. The architecture is the same as used in \cite{b7}. Each convolutional layer is followed by a max pooling. After the third pair of convolution and pooling, one fully connected layer is added, with another one as the output.}
\label{fig:CNN}
\end{figure}

\subsection{IEEE 68-Bus System}\label{sec:CNN_comparison}
The CNN and GNN are trained to classify dataset 1 from the IEEE 68-bus system and the results are compared for varying network sizes. The CNN architecture is the same as the one used in \cite{b7} and similar to that used in \cite{b39}. It is shown in Fig. \ref{fig:CNN}. All convolutional layers and fully connected layers use ReLU activation except for the output layer which uses a sigmoid activation. The input is the 68 X 68 X 3 image containing the power grid data (bus voltages, active and reactive powers). The output is 0 for an unsafe case and 1 for safe cases. The GNN architecture on the other hand is much smaller with only 4 layers. A convolutional layer receives the input which is then connected to 3 fully connected layers. The last fully connected layer serves as the output. For input to the GNN, 3 graph signals are aggregated at node 1 for a chosen aggregation length $K$. Node 1 is chosen due to high centrality. More details on choice of node are given in section \ref{sec_choice_of_node}. The graph signals for aggregation are the bus voltages $V$, net bus active power $P$ and net bus reactive power $Q$. The individual line active and reactive powers are also appended to have input of size $(K+2)\times 3$. 

For both networks, the dataset is split into 75\% for training, 15\% for validation and 10\% for testing. The loss function $\ell$ is the binary cross entropy loss 
\begin{equation}\label{eqn:loss_func}
    \ell = -\frac{1}{M}\sum_{n=1}^M y_m\log(p(y_m))+(1-y_m)\log(1-p(y_m)),
\end{equation}
where $y_m=0$ or 1 is the label for data point $m$ and $p(y_m)\in\{0,1\}$ is the CNN output for data point $m$. $M$ is the total number of data points used for training.
The optimizer used is the Adam optimizer \cite{b29} with a learning rate of 0.0001 and batch size of 128. The CNN is trained for 2500 epochs. The training is done using TensorFlow 2.8 in python with an NVIDIA GeForce RTX 3070 GPU. By varying the number of neurons in each layer of both neural networks, we are able to get varying number of parameters. 

\begin{table}[htbp]
\caption{GNN Results for Varying Aggregation Lengths on the IEEE 68-Bus System.}
\begin{center}
\begin{tabular}{|c|c|c|}
\hline
\textbf{Aggregation length} & \textbf{\# Parameters}& \textbf{Accuracy}\\
\hline
$K$=1 & 85 & 78.21  \\
\hline
$K$=2 & 105 & 80.85 \\
\hline
$K$=3 & 125 & 81.21\\
\hline
$K$=4 & 145 & 81.05 \\
\hline
$K$=5 & 165 & 81.09 \\
\hline
\end{tabular}
\label{tab1:agglength}
\end{center}
\end{table}

The CNN and GNN accuracy are compared for varying parameter sizes. These network sizes are obtained by varying the number of neurons at each of the layers shown for the CNN in Fig. \ref{fig:CNN}. These variations and the number of parameters for each network size are shown in Table \ref{tab1:CNNvsGNN}. Before these results can be compared results, it is important to determine the optimal aggregation length $K$ for the graph. To do this, the graph signals are aggregated at the same node for various aggregation sequences of length $K= 1,2,3,4,5$ and the results obtained for the various aggregation lengths are shown in Table \ref{tab1:agglength}. The results presented are an average of results obtained on repeating the experiments 4 times. The same GNN size is used for all aggregation lengths however the number of parameters increases with the size of input.  From the results in Table \ref{tab1:agglength}, the maximum accuracy is achieved with $K=3$ and increasing the sequence past that does not generate any improvement in the GNN performance. 

With the best aggregation length $K$ for the system determined, the CNN and GNN results are compared. These results obtained are also shown in Table \ref{tab1:CNNvsGNN}. The results show the average and standard deviation of test accuracy from 4 different experiments for the CNN and the GNN. All GNN results shown are for $K=3$ and aggregation at Node 1. From the results obtained, it is seen that the GNN achieves comparable and even better accuracy with fewer parameters than the CNN.

\begin{table*}[htbp]
\caption{CNN vs GNN Results for Various Network Sizes on the IEEE 68-Bus System.}
\begin{center}
\begin{tabular}{|c|c|c|c|c|c|c|c|}
\hline
\multicolumn{4}{|c}{\textbf{CNN}}&\multicolumn{4}{|c|}{\textbf{GNN}} \\
\hline
\textbf{\# Parameters} & \textbf{\# Neurons}& \textbf{Accuracy}& \textbf{Ave Time (min)}& \textbf{\# Parameters} & \textbf{\# Neurons}& \textbf{Accuracy}& \textbf{Ave Time (min)}\\
\hline
4,811 & [4, 8, 8, 5, 1] & 94.32 $\pm$ 1.48 & 87.3 & 3,376 & [10, 20, 50, 1] & 95.70 $\pm$ 1.50 & 11.5\\
\hline
2,863 & [2, 4, 8, 9, 1] & 93.30 $\pm$ 1.20 & 85.2 & 2,321 & [10, 10, 50, 1] & 94.35 $\pm$ 1.28 & 11.3\\
\hline
2,499 & [4, 4, 4, 5, 1] & 85.16 $\pm$ 0.92 & 83.4 & 1,901 & [10, 10, 40, 1] & 92.19 $\pm$ 0.96 & 11.4\\
\hline
1,553 & [2, 4, 4, 4, 1] & 82.47 $\pm$ 1.30 & 81.6 & 1,061 & [10, 10, 20, 1] & 90.87 $\pm$ 0.93 & 11.3\\
\hline
1,089 & [2, 2, 4, 3, 1] & 80.09 $\pm$ 1.83 & 84.2 & 459 & [5, 10, 9, 1] & 84.71 $\pm$ 2.08 & 11.6\\
\hline
857 & [2, 2, 2, 2, 1] & 80.01 $\pm$ 1.26 & 86.6 & 125 & [2, 4, 5, 1] & 81.13 $\pm$ 1.00 & 11.1\\
\hline
\end{tabular}
\label{tab1:CNNvsGNN}
\end{center}
\end{table*}

The CNN and GNN are also compared on training time for 2500 Epochs and networks with similar number of parameters. These results are also shown in Table \ref{tab1:CNNvsGNN}. From the results, the CNN takes more than 7 times the time it takes to train the GNN for all the network sizes and for the same number of Epochs. This is possibly due to the larger input size for the CNN which requires more memory and thus more time to train for each epoch. The GNN is therefore more efficient, getting comparable accuracy in 1/7 the time it takes the CNN to achieve the same.

\subsection{Choice of Aggregation Node}\label{sec_choice_of_node}
\begin{figure}
\centering
\begin{subfigure}{0.4\textwidth}
    \includegraphics[width=\textwidth, height=4.5cm]{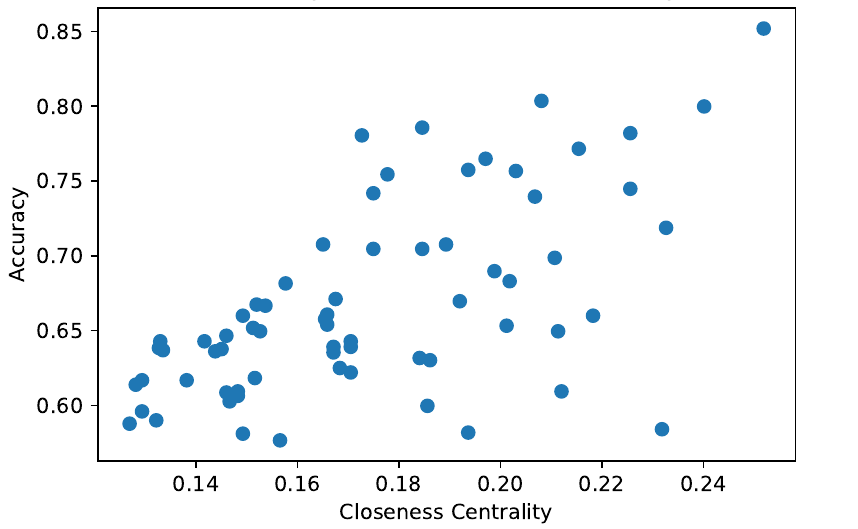}
    \caption{Closeness centrality}
    \label{fig:first}
\end{subfigure}
\hfill
\begin{subfigure}{0.4\textwidth}
    \includegraphics[width=\textwidth, height=4.5cm]{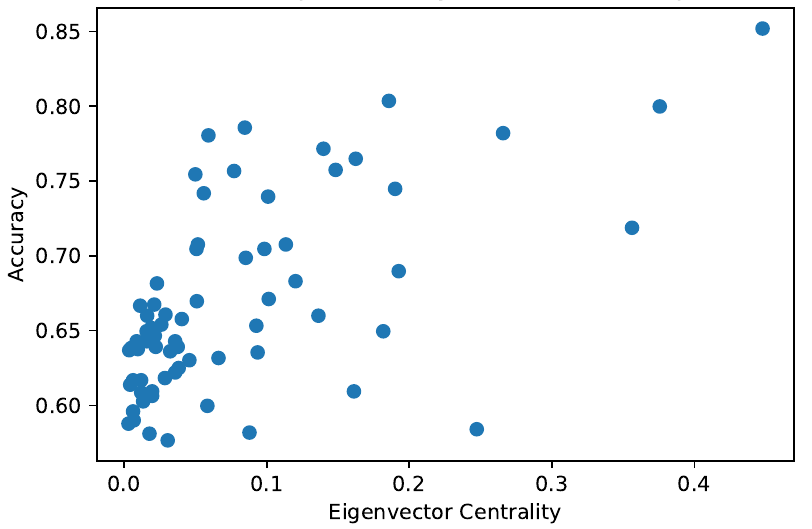}
    \caption{Eigenvector centrality}
    \label{fig:second}
\end{subfigure}        
\caption{Centrality versus GNN accuracy for network of 459 parameters. The closeness centrality shows a correlation coefficient of 0.6158 and eigenvector centrality shows correlation coefficient 0.6128. The node with highest centrality (Node 1) consistently gives maximum accuracy.}
\label{fig:centrality_plot}
\end{figure}

\begin{figure}[htbp]
\centering
\includegraphics[width=9cm, height=6.5cm]{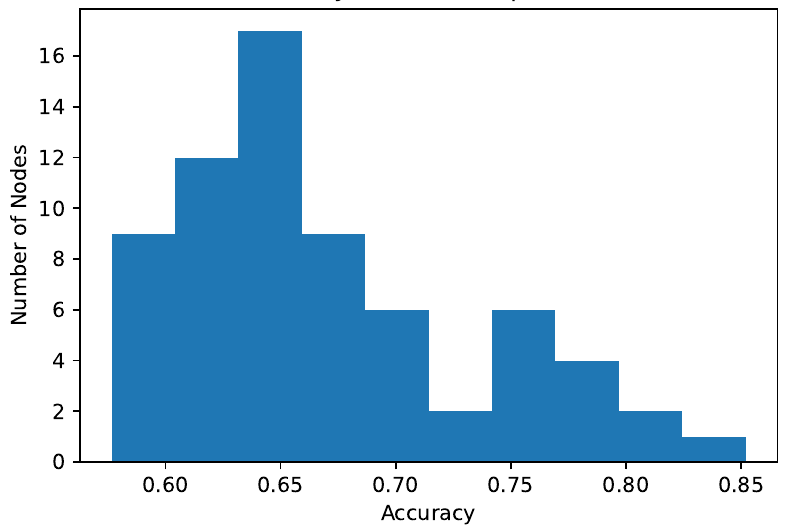}
\caption{Accuracy distribution for the GNN training on all 68 nodes. $K$=3 and the network size is 459 parameters, average of 3 experiments. The distribution shows that only one node gives up to 85\% accuracy.}
\label{fig:Acc_dist}
\end{figure}

All the GNN results shown in Section \ref{sec:CNN_comparison} are for aggregation at Node 1. There are 68 nodes in the network. For the aggregation GNN, it is important to choose the best node for optimal performance. Training on all nodes for the 459 parameter network, 2500 Epochs and $K$=3 gives an average accuracy distribution shown in Fig. \ref{fig:Acc_dist}, after repeating 3 times. From the accuracy distribution, it is seen that only one node results in $85\%$ accuracy. Three nodes result in $80 \%$ accuracy while 1 single node gives the consistent best results for that network size. Training for 2500 Epochs takes on average 11 minutes, with aggregation at only one node. However, training at all 68 nodes will take $68\times11=$ 748 minutes, which is  more time than the CNN. Hence, it is necessary to be able to determine that node of maximum accuracy without training at all nodes and comparing the results. The method proposed in this paper for determining optimal node choice, as stated earlier, is using degree and centrality measures. The closeness centrality (see e.g., \cite{b27}) and eigenvector centrality (see e.g., \cite{b27}) for the IEEE 68 bus network are computed according to (\ref{eqn:close_cen}) and (\ref{eqn:eigen_cen}). These centrality measures are then compared with the accuracy obtained from the distribution in Fig. \ref{fig:Acc_dist}. For the network size of 459 parameters, a positive correlation was obtained between closeness centrality and accuracy, with a correlation coefficient of 0.6158. Eigenvector centrality was also computed with $\lambda=2$. Increasing $\lambda$ showed no remarkable changes in the centrality distribution so $\lambda=2$ was chosen. The eigenvector centrality showed a positive correlation with correlation coefficient 0.6128. These plots of centrality versus accuracy are shown in Fig. \ref{fig:centrality_plot}. Increasing the network size to 1061 parameters also showed a positive correlation with correlation coefficients 0.6666 for closeness centrality and 0.6846 for eigenvector centrality. For every training, Node 1, which is the node with highest degree and centrality, consistently gives maximum accuracy. 

\subsection{NPCC 140-Bus System}\label{sec_multi_node_agg}
For larger networks, aggregation at a single node might not sufficiently capture all the important information in the data. This can be due to associated communication overhead or numerical instabilities \cite{b24}. This can be overcome using multi-node aggregation. This is illustrated using dataset 2 from the NPCC 140 bus system \cite{b28}. This system has 16 generators, 140 buses and 233 transmission lines. This is significantly larger than the IEEE 68 bus system in Section \ref{sec:CNN_comparison}. First the same GNN is trained using aggregation on a single node. The same computer is used as in Section \ref{sec:CNN_comparison}. Training for 2500 Epochs the GNN with the same architecture used previously, generate the results shown in Table \ref{tab1:CNNvsGNN_multi_node}. The results from 4 experiments are shown with the number of parameters in the network, average accuracy, standard deviation and training time for 2500 Epochs. For this system, the aggregation for the GNN is done at Node 37 which is the node of highest closeness centrality and for $K=4$. 

From the results in Table \ref{tab1:CNNvsGNN_multi_node}, it is seen that the performance of the GNN with single node aggregation reduces significantly with the 140-bus system compared to the 68-bus system. This is possibly due to the much larger size of the network. To rectify this, aggregations at multiple nodes are combined. To determine the number of nodes needed, first the graph is divided into communities as done in \cite{b24}. The communities are determined using edge betweenness centrality \cite{b30}. Using this method, 3 communities are obtained. Then the nodes with highest degree and centrality in each community are chosen. These nodes are node 37 for community 1, node 54 for community 2 and node 105 for community 3. Concatenating the aggregation results from all 3 nodes, an input of size $K\times3$ X 3 is obtained. The results after 2500 Epochs of training are also shown in Table \ref{tab1:CNNvsGNN_multi_node}.

Comparing the results with the CNN results in Table \ref{tab1:CNNvsGNN_multi_node}, it is seen that the multi-node GNN outperforms the CNN. The standard deviation is also very low showing consistency in results over the four averaged experiments. The CNN takes on average 83 minutes to train for 2500 Epochs while the multi-node GNN takes 13 minutes on average to train for the same number of epochs. The GNN is thus 6 times faster and more computationally efficient.

\begin{table*}[htbp]
\caption{CNN vs GNN for NPCC 140-Bus System with multi-node aggregation.}
\begin{center}
\begin{tabular}{|c|c|c|c|c|c|c|c|c|}
\hline
\multicolumn{3}{|c}{CNN}&\multicolumn{3}{|c|}{GNN (single node)} &\multicolumn{3}{|c|}{GNN (multi-node)}\\
\hline
\textbf{\# Parameters} & \textbf{Accuracy}& \textbf{Time (min)}&\textbf{\# Parameters} & \textbf{Accuracy} & \textbf{Time (min)}& \textbf{\# Parameters} & \textbf{Accuracy} & \textbf{Time (min)}\\
\hline
10,931 & 94.72 $\pm$ 0.42 & 83.4 &  1,321 & 81.68 $\pm$ 2.54 & 11.4 & 4,216 & 96.72 $\pm$ 0.25 & 13.5\\
\hline
5,559 & 94.57 $\pm$ 0.72 & 83.1 & 1,101 & 81.17 $\pm$ 2.51 & 11.3 & 3,396 & 96.29 $\pm$ 0.52 & 13.3\\
\hline
4,001 & 85.69 $\pm$ 9.60 & 81.7 & 996 & 82.89 $\pm$ 4.14 & 11.4 & 2,576 & 95.04 $\pm$ 0.45 & 13.1\\
\hline
2,925 & 79.50 $\pm$ 11.40 & 82.5 & 776  & 80.00 $\pm$ 2.53 & 11.3 & 1,756 & 96.04 $\pm$ 0.60 & 12.9\\
\hline
1,469 & 71.55 $\pm$ 0.56 & 83.3 & 556 & 80.75 $\pm$ 2.30 & 11.2 & 854 & 95.66 $\pm$ 0.49 & 12.9\\
\hline
834 & 71.19 $\pm$ 1.92 & 82.7 & 314 & 81.29 $\pm$ 2.18 & 11.2 & 205 & 90.98 $\pm$ 2.70 & 12.2 \\
\hline
\end{tabular}
\label{tab1:CNNvsGNN_multi_node}
\end{center}
\end{table*}

\begin{table*}[htbp]
\caption{CNN vs GNN Results With Missing PMU Data.}
\begin{center}
\begin{tabular}{|c||c|c|c|c|}
\hline
 & \textbf{Metric} & \textbf{0 \% Missing Data} & \textbf{10\% Missing Data} & \textbf{20\% Missing Data} \\
\hline
CNN (68 Bus) & $A^c$ & 94.32 & 70.62 & 57.33\\
    & $S^p$ & 93.68 & 69.89 & 96.70\\
    & $R^c$ & 95.15 & 71.56 & 06.96\\
\hline
GNN (68 Bus) & $A^c$ & 95.70 & 95.70 & 95.70 \\
    & $S^p$ & 96.23 & 96.23 & 96.23\\
    & $R^c$ & 95.09 & 95.09 & 95.09\\
\hline
CNN (140 Bus) & $A^c$ & 94.72 & 59.26 & 60.15\\
    & $S^p$ & 95.87 & 79.09 & 86.01\\
    & $R^c$ & 91.67 & 50.23 & 49.40\\
\hline
GNN (140 Bus) & $A^c$ & 96.72 & 96.72 & 96.72 \\
    & $S^p$ & 98.23 & 98.23 & 98.23\\
    & $R^c$ & 99.09 & 99.09 & 99.09\\
\hline
\end{tabular}
\label{tab1:missing_bus}
\end{center}
\end{table*}

\subsection{Robustness to Missing PMU Data}\label{sec_transferability}
In this section, the CNN and GNN are compared on robustness to missing data. The optimal location of the PMUs in both power systems are selected using centrality measures. The results shown so far are for full observability. However, missing data with PMUs and partial observability are unavoidable \cite{b39}. To address this scenario, the GNN and CNN performance is compared for cases where 0-20\% of the buses in the network are unobservable. First the optimal locations for PMU placement are determined using node centrality. These PMU placements are shown in Fig. \ref{fig:68bus} and Fig. \ref{fig:140bus} with the 30\% highest centrality buses highlighted. Next the largest trained CNN and GNN with full observability are tested on cases with 0-20\% missing data at the unhighlighted buses.

To better characterize the results, two extra metrics are included. These are the specificity and recall. These metrics are dependent on the True positives TP (correctly predicted secure conditions), True Negatives TN (correctly predicted unsecure conditions), False Positives FP (unsecure conditions wrongly predicted as secure) and False Negatives FN(secure conditions wrongly predicted as unsecure). The Specificity $S^p$ measures the model's ability to correctly predict unsecure cases while the Recall $R^c$ measures the ability to correctly predict secure cases. They are defined as follows:
\begin{equation}
    R^c = \frac{TP}{TP + FN}
\end{equation}
\begin{equation}
    S^p = \frac{TN}{TN + FP}
\end{equation}
These results are summarized in Table \ref{tab1:missing_bus} with $A^c$ representing the accuracy. From the results, the GNN significantly outperforms the CNN showing robustness with no change in accuracy for up to 20\% missing data. 

\subsection{Online Efficiency}\label{sec_testing_time}
Finally, the GNN is analyzed on testing efficiency for online application. The time needed to test one batch of data is compared for time domain simulations, the CNN and the GNN. These results are summarized in Table \ref{tab1:testing_time}. From the results, the GNN is the most efficient requiring around 10\% the time required by the CNN.

\section{Conclusion}\label{sec_conclusion}
This paper proposes a graph neural network approach to SSA. Leveraging the graphical nature of a power grid, we apply GNNs, to improve stability and robustness. From the results in Sections \ref{sec:CNN_comparison} and \ref{sec_multi_node_agg}, we see that the GNN is able to perform more efficiently than the CNN, requiring a very small number of parameters to train and successfully classify the dataset with over 95\% accuracy. We also see that the trained GNN displays significant robustness over the trained CNN  with missing data at up to 20\% of buses in the network. This shows that GNNs can successfully exploit the graphical nature of the grid to improve stability, efficiency and robustness.

\begin{table}[htbp]
\caption{Average Execution Time for Real-Time Application}
\begin{center}
\begin{tabular}{|c|c|c|c|}
\hline
 & \textbf{Time Domain Simulations} &\textbf{CNN}&\textbf{GNN} \\
\hline
140 Bus & 629s & 30ms & 4ms \\
\hline
68 Bus & 215s & 19ms & 3ms \\
\hline
\end{tabular}
\label{tab1:testing_time}
\end{center}
\end{table}

\noindent\textbf{Glory Justin}  received the Bachelor of Engineering degree in electrical and electronic engineering from Federal University of Technology Owerri, Imo State, Nigeria, in 2017 and the M.S. degree in electrical engineering from Rensselaer Polytechnic Institute, Troy New York, in 2022. She has done summer internships with Heliogen as an Engineering Intern in 2022 and GE Aerospace research as an Electric Machines and High Voltage Engineering Intern in 2023. She is currently pursuing the Ph.D. degree in electrical engineering at Rensselaer Polytechnic Institute, Troy, NY. Her research interests include power system stability and control, and machine learning.\\

\noindent\textbf{Santiago Paternain} received the B.Sc. degree in electrical engineering from Universidad de la República Oriental del Uruguay, Montevideo, Uruguay in 2012, the M.Sc. in Statistics from the Wharton School in 2018 and the Ph.D. in Electrical and Systems Engineering from the Department of Electrical and Systems Engineering, the University of Pennsylvania in 2018. He is currently an Assistant Professor in the Department of Electrical Computer and Systems Engineering at the Rensselaer Polytechnic Institute. Prior to joining Rensselaer, he was a postdoctoral Researcher at the University of Pennsylvania. He was the recipient of the 2017 CDC Best Student Paper Award and the 2019 Joseph and Rosaline Wolfe Best Doctoral Dissertation Award from the Electrical and Systems Engineering Department at the University of Pennsylvania. His research interests lie at the intersection of machine learning and control of dynamical systems.\\

\end{document}